\documentclass[a4paper,10pt]{article}
\usepackage[latin9]{inputenc}

\usepackage{amsmath}
\usepackage{amssymb}
\usepackage{amsthm}
\usepackage{mathtools}

\theoremstyle{plain}
\newtheorem{theorem}{Theorem}

\newtheorem{myrule}{Rule}

\theoremstyle{definition}
\newtheorem{definition}[theorem]{Definition}
\newtheorem{example}[theorem]{Example}

\usepackage{setspace}
\usepackage[hidelinks]{hyperref}

\usepackage{enumitem}
\usepackage{qtree}

\usepackage{framed}

\newcommand{\incmp}{\not\gtrless}

\usepackage[british]{babel}
\author{Luc Edixhoven \and Walter Kosters}
\title{Player preferences in $N$-player games}

\begin{document}

\maketitle

\begin{abstract}
In this paper we describe several player preferences in games with $N \geq 2$ players, in particular the case $N = 3$, and use them to simplify game trees, using the game of Clobber as our example. We show that, using a fixed starting player and a certain ruleset, any short game can be simplified to a value in a very concise set. Omitting the fixed starting player and generalising the theory to more than 3 players remains a challenge.
\end{abstract}

\section{Introduction}
In the words of Aaron Siegel, and we could not have said it better ourselves:

\hspace{6pt}\textit{Combinatorial game theory is the study of two-player games with no hidden information and no chance elements. The theory assigns algebraic values to positions in such games and seeks to quantify the algebraic and combinatorial structure of their interactions.}

\smallskip

Combinatorial Game Theory by Siegel~\cite{cgt} and Winning Ways for Your Mathematical Plays by Berlekamp, Conway and Guy~\cite{winningways} are the foremost literature on combinatorial game theory. While the classical theory only considers two-player games, in recent years, more and more work has been done to extend the theory to games with more than two players, both for specifically three and for an arbitrary finite number of players. Most of these efforts make restrictive assumptions about the behaviour of the players, as mentioned by Cincotti~\cite{cincotti}.
Cincotti presents a theoretical framework to classify partisan games with an arbitrary finite number of players. We decided to see how far we could simplify the game tree using as few assumptions as possible, using the game of Clobber as an example to apply our theory.

In Section~\ref{sec:N-player games} we define the games we study in this paper and the notation we use to represent their values. Section~\ref{sec:Clobber} introduces the game of Clobber and how to extend it to an $N$-player game. In Section~\ref{sec:simplifying} we present our simplification rules, the results of which we discuss in Section~\ref{sec:results}. Finally, we summarise our findings and conclusions in Section~\ref{sec:conclusions}, in which we also suggest several areas for further research.

This research was done by the first author as a Master research project at the Leiden Institute of Advanced Computer Science (LIACS), Leiden University, under the supervision of the second author.

\section{N-player games}
\label{sec:N-player games}
We consider a game with $N \geq 2$ players, where we are most interested in the case $N > 2$. The players are numbered $1,2,\ldots,N$. They take turns, where player $i + 1$ succeeds player $i$ (if $i \in \{1,2,\ldots,N-1\}$), and player $1$ succeeds player $N$. Player $1$ starts the game. The last player that can make a legal move, wins the game. This is called \emph{normal play}. If a player cannot make a valid move, their turn is skipped. We assume that at least one player can make a move in the initial position. Furthermore, we assume the game to be \emph{converging}: positions can be ordered in a game tree without backlinks.

We first construct the full game tree, starting from the initial position. Leaves are positions where no player can move. The value of such a node is equal to the number of the winner. These values represent unconditional wins for the corresponding player.

Now we can recursively label all nodes, in the following bottom-up way. The general value of a non-leaf position $P$ with player $i$ to play (note that $i$ is formally part of $P$, and could therefore be omitted) is the list $L$ with all unique values of the children, using some fixed ordering. The underlying intuition is that the list elements represent the choices for the player to move. A value $L$ thus represents a tree, with the leaves having the aforementioned single number values. $L$ is said to \emph{contain} a value $a$ if some node in the tree represented by $L$ has the value $a$.

Note that if all children have the same value, this will also be the value of the parent. We identify a list $[x]$ with $x$ a leaf position with its only member $x$: we use $3$ instead of $[3]$. However, note that, e.g., $[1,3]$ differs from $[[1,3]]$. Here, the first list denotes a situation where the player to move can select $1$ or $3$ as the winner, whereas the second list passes this option to the next player to move. But $[[[3]]] = 3$.

Of course, the order of the list elements does not matter and multiple occurrences of elements can be represented with single occurrences.

We mention some examples:

\begin{framed}
\begin{example}
Suppose the children have values 2, 2, 2 and 3, respectively; then the parent has value $[2,3]$. The parent contains the values $2$, $3$ and $[2,3]$.

Note that, if it is player 1's turn, this value makes player 1 a so-called \emph{kingmaker}. As also noted by Propp~\cite{propp}, the player has no winning move, but their action determines which of the other players will win.
\end{example}
\begin{example}
Suppose the children have values 2, $[2,3]$, $[2,3]$ and $[1,[1,3]]$, respectively; then the parent has value $[2,[2,3],[1,[1,3]]]$. The parent contains the values $1$, $2$, $3$, $[1,3]$, $[2,3]$, $[1,[1,3]]$ and $[2,[2,3],[1,[1,3]]]$.
\end{example}
\end{framed}

\section{Clobber}
\label{sec:Clobber}
Clobber is a \emph{partisan game} consisting of an undirected graph, usually a grid graph, with the vertices containing a black or white token or being empty. A player must move one of their tokens to an adjacent vertex containing a token of the opponent. The player's token replaces, ``clobbers'', the opponent's token, which is then removed from the game. The first player unable to make such a move loses the game. Note that Clobber is \emph{dicotic}, formally known as \emph{all-small}~\cite[pages 60--63]{cgt}, meaning that both players can move from every nonempty position. In competitions, Clobber is usually played on a checkerboard with black tokens on the black squares and white tokens on the white ones. Human competitions usually use a $5 \times 6$ board while computer competitions generally use larger board sizes such as $10 \times 10$.

For further reading on Clobber, we recommend the 2005 paper ``An introduction to Clobber''~\cite{introductiontoclobber} and Siegel's 2013 book on combinatorial game theory~\cite[pages 146--149]{cgt}. Recent work on Clobber was done in 2016 by Griebel and Uiterwijk~\cite{griebeluiterwijk}, who combined combinatorial game theory with an $\alpha$-$\beta$-solver to solve larger and more complex Clobber boards.

To extend Clobber into an $N$-player game, a vertex now contains a number between 0 and $N$, 0 meaning the vertex is empty and a number $i \geq 1$ meaning the vertex contains a token from the corresponding player $i$. A valid move now consists of clobbering an adjacent token belonging to any other player. As defined in Section~\ref{sec:N-player games}, a player unable to make a valid move will skip their turn --- and can never move again --- and the last player to make a valid move wins the game.

We now give some examples of three-player Clobber games on $1 \times n$ boards and their values. In all examples, we assume it is player 1's turn.

\begin{framed}

\begin{example}
\begin{tabular}{| c | c | c |}
\hline
2 & 1 & 3 \\
\hline
\end{tabular}
has value 1. Player 1 can clobber either player 2 or 3 and wins in both cases.
\end{example}

\begin{example}
\begin{tabular}{| c | c | c | c | c |}
\hline
1 & 2 & 2 & 2 & 3 \\
\hline
\end{tabular}
has value [[1,3]]. Player 1 has no choice but to clobber player 2. Player 2 then must clobber either player 1 or 3, after which the other one will clobber them in return and win. Player 2 thus chooses the winner.
\end{example}

\begin{example}
\begin{tabular}{| c | c | c | c | c | c |}
\hline
1 & 2 & 3 & 2 & 1 & 3 \\
\hline
\end{tabular}
has value [[1,3],[1,[1,2]],[2,3]]. This can still easily be checked by hand, which we leave as an exercise for the reader.
\end{example}

\begin{example}
\label{example:1x10-unsimplified}
\begin{tabular}{| c | c | c | c | c | c | c | c | c | c |}
\hline
1 & 2 & 3 & 2 & 1 & 3 & 2 & 3 & 2 & 1 \\
\hline
\end{tabular}
has value [[[1, 3, [3, [[1, 3]]], [[1, 3, [2, 3]], [2, 3, [1, 2]]], [[[1, 2]]]], [3, [2, 3], [2, [1, 3]], [[1, 2]]], [3, [3, [1, 3]], [3, [[1, 3]]], [[1, 3], [2, 3]], [[1, 3]]], [3, [[1, 2], [2, 3]]], [[1, 3], [3, [1, 2], [[1, 2]]], [[1, 2, 3], [1, 2, [2, 3]]]], [[1, 3], [3, [1, 2], [[1, 2]]], [[1, 3], [2, 3, [2, 3]], [2, 3]]]], [[2, [1, 3], [2, 3], [[1, 3]]], [2, [3, [1, 3]]], [3, [1, [2, 3]], [[1, 3, [2, 3]]], [[1, 3], [[1, 2]]], [[2, 3], [[1, 3]]]], [[1, 3], [1, [2, 3]], [3, [2, 3]], [[2, 3]]], [[1, [1, 2, 3], [2, 3, [2, 3]]], [2, 3], [2, [[1, 2]]]], [[2, 3], [2, [1, 2]], [[1, 2, [2, 3]], [1, 2]]]], [[2, [2, 3]], [2, [3, [1, 3]], [3, [2, 3]]], [2, [3, [1, 3]]], [3, [2, 3], [[1, 3]], [[2, 3], [[1, 3]]]], [[1, 2], [1, 3, [1, 3]], [2, 3]], [[1, 2], [2, 3]]], [[[1, 2], [1, [2, 3]], [2, 3]], [[1, 2], [2, 3]], [[1, 3, [1, 2]]], [[2, 3]], [[3, [1, 3]], [[1, 3]]]]]. Some spaces added for readability. We do not recommend to check this one without the assistance of a computer.
\end{example}

\end{framed}

\section{Simplifying the game tree}
\label{sec:simplifying}
We have seen from the examples in Section~\ref{sec:Clobber} that the length and complexity of values grow rather fiercely for larger board sizes. To counter this, for games with $N = 3$, we introduce an additional notation, give several general, syntactic simplification rules and experiment with several player preferences and the semantic simplification rules they infer. Note that these do not rely on any rules specific to Clobber and should instead be applicable to any short game, as defined in~\cite[page 54]{cgt}.

\subsection{Simple values}
\label{sec:simplevalues}
For $N = 3$, we use $\bar{1}$ (pronounced ``1 bar'') to denote the value $[2,3]$. $\bar{1}$ can be interpreted as the complement of 1, as it consists of all single number values except for 1 itself.
Similarly, we use $\bar{2}$ for $[1,3]$ and $\bar{3}$ for $[1,2]$.
Following the same intuition, we use the notation $\bar{\bar{1}}$ (pronounced ``1 bar bar'') for $[\bar{2},\bar{3}]~(\left[[1,3],[1,2]\right])$, and so forth.
Due to the large number of bars in larger and more complex values, we will often omit the actual bars and instead denote their number in subscript; for instance, $1_2 = \bar{\bar{1}}$ and $1_6 = \bar{\bar{\bar{\bar{\bar{\bar{1}}}}}}$.
We use the notation $a_i$, $a$ being the \emph{base value} and $i \geq 0$ being the \emph{exponent}, to denote a value consisting of the list containing the values $\{b_{i-1}\,|\,b \neq a\}$, with $a_0$ representing $a$, the unconditional win for player $a$. We call values that can be represented using this notation \emph{simple values}. We call all other values \emph{complex values}.

\subsection{Syntactic simplifications}
\label{sec:syntaxsimpl}
We give several operations, denoted by ``$\Rightarrow$'', that can be performed to simplify the syntax of the game tree without semantically changing the possibilities available to the players or the possible outcomes of the game tree.

\begin{framed}
\begin{myrule}
\label{rule:simplebrackets}
For a simple value $x$: $[x] \Rightarrow x$.
\end{myrule}
In our previous notation, there was a semantic difference between, e.g., $[1,3]$ and $[[1,3]]$, as a different player makes the choice between the values 1 and 3. However, a simple value encapsulates this as its base value is the player making the choice, e.g. $\bar{2}$ can be interpreted as ``Regardless of what else happens, at some moment player 2 will make a choice between two moves leading to positions with values 1 and 3.'' Therefore, $\bar{2}$ can be used to represent both $[1,3]$ and $[[1,3]]$.

In general, a simple value defines a complete binary game tree, which contains all possible outcomes, the choices leading to these outcomes and which player makes which choice. Of course, the value $\bar{2}$ could represent a game where a hundred moves are played without any choice being involved, or where every choice leads to positions with the same values, before player 2 makes their deciding choice, and several hundred more moves could be played after this choice, but this does not change the outcome.
\end{framed}

\begin{framed}
\begin{myrule}
\label{rule:triplebrackets}
For a (simple or complex) value $x$ with $N = 3$: $[[[x]]] \Rightarrow x$.

Or, in general, with $N$ the number of players: $x = \underbrace{[[[\ldots[[[}_N x ]]]\ldots]]]$.
\vspace{-6pt}
\end{myrule}
This rule again uses the argument that we can omit nodes of the game tree if they do not influence the possible outcomes or the choices leading to them in any way. Two values $x$ and $[[[x]]]$ are only different in that the latter has three extra moves leading up to the same choice. As these moves do not influence the choice or its outcomes, and the player who is to make these choices is the same, the values are semantically equivalent and we can omit the three sets of square brackets.
\end{framed}

\begin{framed}
\begin{myrule}
\label{rule:sameplayerchoice}
For (simple or complex) values $x_1,\ldots,x_m$ and $y_1,\ldots,y_k$ with $N = 3$: \\
$[x_1,\ldots,x_m,[ [ [y_1,\ldots,y_k]]]] \Rightarrow [x_1,\ldots,x_m,y_1,\ldots,y_k]$.

Or, in general, with $N$ the number of players: \\
$[x_1,\ldots,x_m,\underbrace{[[[\ldots[[[}_Ny_1,\ldots,y_k]]]\ldots]]]] \Rightarrow [x_1,\ldots,x_m,y_1,\ldots,y_k]$
\vspace{-6pt}
\end{myrule}
This rule can be used together with Rule~\ref{rule:triplebrackets} to merge nodes with their ancestors in the game tree if no choices by other players were involved on the path between them. This builds upon the intuition used in Rule~\ref{rule:triplebrackets} that, as long as the intermediate nodes where other players can make a move do not branch, the same player keeps making every choice, giving them complete control over the possible outcomes. The outcomes can thus be merged into the children of a single node, a single list, as this does not reduce the possibilities available to the players.

For instance, this rule can be used to simplify $[\bar{\bar{1}},\bar{2}]$ into $\bar{\bar{1}}$, assuming it is player 1's turn. As $\bar{\bar{1}} = [\bar{2},\bar{3}]$, player 1 has the choice between choosing for $\bar{2}$ immediately or taking a different path in which they will eventually choose between $\bar{2}$ and $\bar{3}$. In the end, player 1 chooses between $\bar{2}$ and $\bar{3}$ without any influence from the other players, so the original choice can be simplified to $[\bar{2},\bar{3}] = \bar{\bar{1}}$.
\end{framed}

\subsection{Player preferences}
\label{sec:preferences}
Until now, we have only considered simplification rules that do not actually change the semantics of the game tree. We now consider possible player preferences to actually discard certain values and thus prune the game tree. We define a binary relation over the values of game trees: a value $X$ is said to be \emph{weaker than or equal to} a value $Y$ from the perspective of player $p$, written as $X \leq_p Y$, if they represent the same value or if the relation can be inferred from the values of their children. Formally, with $X = [x_1,\dotsc,x_n], Y = [y_1,\dotsc,y_m]$, we define the relation $\leq_p$ as follows:

\begin{framed}
\begin{definition}
\label{def:leqp}
	$X \leq_p Y \Leftrightarrow (X = Y) \text{ or } (\forall i: x_i \leq_p Y) \text{ or } (\forall i,j: x_i \leq_p y_j)$
\end{definition}

With $X = Y$, we mean that $X$ and $Y$ are exactly the same values after using syntactic simplifications as in Section~\ref{sec:syntaxsimpl}.
\end{framed}

Using this definition, we can define three more relations:

Two values $X$ and $Y$ are \emph{equal} to each other from the perspective of player $p$, written as $X =_p Y$, if they are both weaker than or equal to each other. Formally:
\begin{framed}
\begin{definition}
\label{def:eqp}
$X =_p Y \Leftrightarrow X \leq_p Y \text{ and } Y \leq_p X$
\end{definition}
\end{framed}
A value $X$ is \emph{strictly weaker} than a value $Y$ from the perspective of player $p$, written as $X <_p Y$, if $X$ is weaker than or equal to $Y$ and they are not equal. Formally:
\begin{framed}
\begin{definition}
\label{def:ltp}
$X <_p Y \Leftrightarrow X \leq_p Y \text{ and } X \neq_p Y$
\end{definition}
\end{framed}
A value $X$ is \emph{incomparable} with a value $Y$ from the perspective of player $p$, written as $X \incmp_p Y$, if $X$ is not weaker than or equal to $Y$ and $Y$ is not weaker than or equal to $X$. Formally:
\begin{framed}
\begin{definition}
\label{def:incmpp}
$X \incmp_p Y \Leftrightarrow X \nleq_p Y \text{ and } Y \nleq_p X$
\end{definition}
\end{framed}

\newpage

\subsection{Selfish play}
\label{sec:selfish}
A logical first player preference to introduce would be a \emph{selfish} player --- when faced with a choice between two values, they will choose the value that is in their own best interest. In particular, this implies that the player will always choose a move where they will certainly win. If no such move is present, they will choose a move where they might win. Formally, it allows us to define two additional rules. We use the term \emph{selfish game} to denote a game where all players are selfish.

\begin{framed}
\begin{myrule}
\label{rule:guaranteedwin}
Assuming a selfish player $a$, it being player $a$'s turn, and $x$ being an arbitrary value that is neither a guaranteed win or loss, then: $x <_a a$.
\end{myrule}
If player $a$ can make a move leading to unconditional victory, they will choose to do so and disregard all other moves.
\end{framed}

\begin{framed}
\begin{myrule}
\label{rule:avoidloss}
Assuming a selfish player $a$, it being player $a$'s turn, $x$ being an arbitrary value that is neither a guaranteed win or loss, and $y$ being an arbitrary value that is a guaranteed loss, then: $y <_a x$.
\end{myrule}
As player $a$ plays to win and the value $y$ and $c$ represents a guaranteed loss, the player will prefer any value $x$ that still has some possibility, no matter how small, to lead to a victory.
\end{framed}

We can use the above rules to show that $\bar{\bar{3}} <_1 \bar{2}$. After all, $\bar{\bar{3}} = [\bar{1},\bar{2}]$. Rule~\ref{rule:avoidloss} gives us that $\bar{1} <_1 \bar{2}$. We can conclude from this that $\bar{\bar{3}} <_1 \bar{2}$. Note that using this and Definition~\ref{def:leqp}, we can also show that $[\bar{2},\bar{\bar{3}}] <_1 \bar{2}$, and then that $[\bar{2},[\bar{2},\bar{\bar{3}}]] <_1 \bar{2}$, and so on. However, $\bar{2}$ and $\bar{\bar{2}}$ for instance are incomparable. Since we have no way of comparing 2 and 3 from the perspective of player 1, we also have no way of comparing $\bar{2}$ and $\bar{3}$ and so forth. This significantly limits the gains of the simplification rules so far. Furthermore, we would actually like to be able to compare $\bar{2}$ and $\bar{\bar{2}}$; both have a single path where player 1 wins, but $\bar{\bar{2}}$ has three paths where player 1 loses compared to a single path in $\bar{2}$, and player 1 has no way to steer towards its winning path in either case. Although we assume nothing about the preferences of players 2 and 3 beyond them playing selfishly, it would seem wise to prefer $\bar{2}$ over $\bar{\bar{2}}$ as player 1. Therefore, we need something more.

\subsection{Prudently selfish play}
\label{sec:prudent}

To counter the issue we raised at the end of Section~\ref{sec:selfish}, we introduce the notion of a \emph{prudently selfish} player, or a \emph{prudent} player for short --- and \emph{prudent game} for a game with only prudent players. A prudent player will, in addition to playing selfishly, when choosing between two values $X$ and $Y$ where neither is strictly weaker than the other, avoid one if it can lead to a situation that is worse than or incomparable to every single situation in the other one. If such a choice occurs where the option $X$ is discarded in favour of $Y$, we say that, from the point of view of player $p$, $X$ is \emph{prudently weaker} than $Y$, written as $X <^P_p Y$. Formally, with $X = [x_1,\dotsc,x_n], Y = [y_1,\dotsc,y_m]$, we define the relation $<^P_p$ and its corresponding incomparability relation $\incmp^P_p$ as follows:

\begin{framed}
\begin{definition}
$X <^P_p Y \Leftrightarrow X <_p Y \text{ or } (\forall i,j: (x_i <^P_p y_j \text{ or } x_i \incmp^P_p y_j) \text{ and } \exists i,j: (x_i <^P_p y_j))$
\end{definition}
\end{framed}
\begin{framed}
\begin{definition}
$X \incmp^P_p Y \Leftrightarrow X \nless^P_p Y \text{ and } Y \nless^P_p X$
\end{definition}
\end{framed}

We now show that this new relation $<^P_p$ allows us to compare almost every single pair of simple values and that we can use it to simplify any complex value to a single simple value. To this end, we prove with three theorems that this holds for the relation $<^P_1$, from the perspective of player $1$. The proofs for the relations $<^P_2$ and $<^P_3$ are the same.

First we prove that the values $2_i$ and $3_i$ are incomparable and that $1_i$ is comparable with $2_i$ and $3_i$, with the sign depending on the parity of $i$:

\begin{framed}
\begin{theorem}
\[
\forall i \geq 0: 2_i \incmp^P_1 3_i \land
\begin{cases*}
2_i <^P_1 1_i \land 3_i <^P_1 1_i & when $i$ is even \\
1_i <^P_1 2_i \land 1_i <^P_1 3_i & when $i$ is odd \\
\end{cases*}
\]
\end{theorem}

We prove this by induction. Our base case is $i = 0$: $2 \incmp^P_1 3$. It can easily be determined that $2$ and $3$ are not comparable under the rules we have defined. Furthermore, we determine that $2 <^P_1 1$ and $3 <^P_1 1$. Both of these can be easily checked.

For our induction step, we assume our hypothesis to hold for all $0 \leq i \leq k$. We now prove this to induce that the hypothesis also holds for $k+1$. Consider the game trees for the values $1_{k+1}$, $2_{k+1}$ and $3_{k+1}$:

\Tree [.$1_{k+1}$ $2_k$ $3_k$ ]
\Tree [.$2_{k+1}$ $1_k$ $3_k$ ]
\Tree [.$3_{k+1}$ $1_k$ $2_k$ ]

We now attempt to compare $2_{k+1}$ with $3_{k+1}$. There are two ways for the two values to be comparable: $2_{k+1} <^P_1 3_{k+1}$ or $3_{k+1} <^P_1 2_{k+1}$. From our induction hypothesis, it follows that $2_k \incmp^P_1 3_k$. What remains is to compare $1_k$ with $2_k$ and $3_k$ with $1_k$. We first consider the option $2_{k+1} <^P_1 3_{k+1}$. Since we know from our induction hypothesis that $1_k$ is comparable both with $2_k$ and $3_k$, we know that both $1_k <^P_1 2_k$ and $3_k <^P_1 1_k$ must hold. This is impossible by our induction hypothesis. The second option, $3_{k+1} <^P_1 2_{k+1}$, can be analogously proven to be impossible. We conclude that $2_{k+1} \incmp^P_1 3_{k+1}$.

We then compare $1_{k+1}$ with $2_{k+1}$. It follows from our induction hypothesis that $2_k \incmp^P_1 3_k$. Furthermore, from hypothesis we know that either $1_k <^P_1 2_k \land 1_k <^P_1 3_k$ or $2_k <^P_1 1_k \land 3_k <^P_1 1_k$. In the first case, when $k$ is even, it follows that $2_{k+1} <^P_1 1_{k+1}$. In a similar fashion it then holds that $3_{k+1} <^P_1 1_{k+1}$. In the second case, when $k$ is odd, the inverse holds: $1_{k+1} <^P_1 2_{k+1} \land 1_{k+1} <^P_1 3_{k+1}$.

Together, these comparisons show that our hypothesis also holds for $k+1$. By mathematical induction, the statement holds for all $i \geq 0$.
\end{framed}

Second we prove that the value $1_{i+1}$ is incomparable with the values ${2_i}$ and ${3_i}$:

\begin{framed}
\begin{theorem}
$\forall i \geq 0: 1_{i+1} \incmp^P_1 2_i \land 1_{i+1} \incmp^P_1 3_i$
\end{theorem}

For this, again consider the tree for $1_{i+1}$:

\Tree [.$1_{i+1}$ $2_i$ $3_i$ ]

From our previous proof, it holds that $2_i \incmp^P_1 3_i$. It follows that there exists no child value of $1_{i+1}$ which is prudently weaker than a child value of $2_i$ and vice versa. Thus, we conclude that $1_{i+1} \incmp^P_1 2_i$. Analogously, it holds that $1_{i+1} \incmp^P_1 3_i$.
\end{framed}

Finally, we prove that the following ordering holds:

\begin{framed}
\begin{theorem}
\label{theorem:prudentordering}
$\{\bar{1},2,3\} <^P_1 \{\bar{\bar{\bar{1}}},\bar{\bar{2}},\bar{\bar{3}}\} <^P_1 \{\bar{\bar{\bar{\bar{\bar{1}}}}},\bar{\bar{\bar{\bar{2}}}},\bar{\bar{\bar{\bar{3}}}}\} <^P_1 \ldots <^P_1 \{\bar{\bar{\bar{\bar{1}}}},\bar{\bar{\bar{2}}},\bar{\bar{\bar{3}}}\} <^P_1 \{\bar{\bar{1}},\bar{2},\bar{3}\} <^P_1 1$ where values within sets of brackets are incomparable with each other.
\end{theorem}

We prove this by induction. Our base case consists of the ordering of the values $1$, $\bar{1}$, $2$, $\bar{2}$, $3$ and $\bar{3}$. Since $1$ is a guaranteed victory and $\bar{1}$, $2$ and $3$ are guaranteed losses, it is easy to place them in the ordering. We know already that $\bar{1}$, $2$ and $3$ are incomparable. Since $\bar{2}$ and $\bar{3}$ are neither guaranteed victories or guaranteed losses, they are weaker than $1$ and stronger than $\bar{1}$, $2$ and $3$. We know already that they are incomparable with each other. Combining all this, we obtain the following ordering:

\begin{center}
$\{\bar{1},2,3\} <^P_1 \{\bar{2},\bar{3}\} <^P_1 1$
\end{center}

For our induction step, we assume the following ordering to hold for some $i > 1$:

\begin{center}
$\{1_1,2,3\} <^P_1 \{1_3,2_2,3_2\} <^P_1 \ldots <^P_1 \{1_{2i-1},2_{2i-2},3_{2i-2}\} <^P_1 \{2_{2i-1},3_{2i-1}\} <^P_1 \{1_{2i-2},2_{2i-3},3_{2i-3}\} <^P_1 \ldots <^P_1 \{1_2,2_1,3_1\} <^P_1 1_0$
\end{center}

We want to prove that the next values, $1_{2i}$, $1_{2i+1}$, $2_{2i}$, $2_{2i+1}$, $3_{2i}$ and $3_{2i+1}$, are inserted in the ordering as follows:

\begin{center}
$\ldots <^P_1 \{1_{2i-1},2_{2i-2},3_{2i-2}\} <^P_1 \mathbf{\{1_{2i+1},2_{2i},3_{2i}\}} <^P_1 \mathbf{\{2_{2i+1},3_{2i+1}\}} <^P_1 \{\mathbf{1_{2i}},2_{2i-1},3_{2i-1}\} <^P_1 \{1_{2i-2},2_{2i-3},3_{2i-3}\} <^P_1 \ldots$
\end{center}

We prove this in four steps:

\begin{enumerate}[label=\textbf{\Roman*)}]

	\item
	\Tree [.$1_{2i}$ $2_{2i-1}$ $3_{2i-1}$ ]

	We know by induction hypothesis that $\{2_{2i-1},3_{2i-1}\} <^P_1 1_{2i-2}$. Therefore, $1_{2i} <^P_1 1_{2i-2}$. We know by induction hypothesis that $\{2_{2i-1},3_{2i-1}\} <^P_1 \{2_{2i-3},3_{2i-3}\}$. Therefore, $1_{2i} <^P_1 \{2_{2i-3},3_{2i-3}\}$. We know from our induction hypothesis that $\{2_{2i-1},3_{2i-1}\} <^P_1 \{1_{2i-2},2_{2i-3},3_{2i-3}\}$ and we know from a previous proof that $1_{2i} \incmp^P_1 \{2_{2i-1},3_{2i-1}\}$.

	Combining the above gives us the ordering $\{\mathbf{1_{2i}},2_{2i-1},3_{2i-1}\} <^P_1 \{1_{2i-2},2_{2i-3},3_{2i-3}\}$.

	\item
	\Tree [.$2_{2i+1}$ $1_{2i}$ [.$3_{2i}$ $1_{2i-1}$ $2_{2i-1}$ ] ]
	\Tree [.$3_{2i+1}$ $1_{2i}$ [.$2_{2i}$ $1_{2i-1}$ $3_{2i-1}$ ] ]

	We know already that $\{1_{2i},2_{2i-1},3_{2i-1}\} \incmp^P_1 \{2_{2i-1},3_{2i-1}\}$, and by induction hypothesis that $\{1_{2i-1}\} <^P_1 \{2_{2i-1},3_{2i-1}\}$. It follows that $\{2_{2i},3_{2i}\} <^P_1 \{2_{2i-1},3_{2i-1}\}$ and thus that $\{2_{2i+1},3_{2i+1}\} <^P_1 \{2_{2i-1},3_{2i-1}\}$. We know from a previous proof that $\{2_{2i},3_{2i}\} <^P_1 1_{2i}$. Therefore, $\{2_{2i+1},3_{2i+1}\} <^P_1 1_{2i}$. We know from a previous proof that $1_{2i} \incmp^P_1 \{2_{2i-1},3_{2i-1}\}$.

	Combining the above gives us the ordering $\mathbf{\{2_{2i+1},3_{2i+1}\}} <^P_1 \{\mathbf{1_{2i}},2_{2i-1},3_{2i-1}\}$.

	\item
	\Tree [.$2_{2i+1}$ $1_{2i}$ $3_{2i}$ ]
	\Tree [.$3_{2i+1}$ $1_{2i}$ $2_{2i}$ ]

	We know from a previous proof that $\{2_{2i},3_{2i}\} <^P_1 \{1_{2i}\}$. It follows that $\{2_{2i},3_{2i}\} <^P_1 \{2_{2i+1},3_{2i+1}\}$. We know from a previous proof that $1_{2i+1} <^P_1 \{2_{2i+1},3_{2i+1}\}$ and $1_{2i+1} \incmp^P_1 \{2_{2i},3_{2i}\}$.

	Combining the above gives us the ordering $\mathbf{\{1_{2i+1},2_{2i},3_{2i}\}} <^P_1 \mathbf{\{2_{2i+1},3_{2i+1}\}}$.

	\item
	\Tree [.$1_{2i+1}$ $2_{2i}$ $3_{2i}$ ]
	\Tree [.$2_{2i}$ $1_{2i-1}$ $3_{2i-1}$ ]
	\Tree [.$3_{2i}$ $1_{2i-1}$ $2_{2i-1}$ ]

	We know from our induction hypothesis that $1_{2i-1} <^P_1 \{2_{2i-1},3_{2i-1}\}$. Therefore, $1_{2i-1} <^P_1 \{2_{2i},3_{2i}\}$, which lets us conclude that $1_{2i-1} <^P_1 1_{2i+1}$. We know from our induction hypothesis that $1_{2i-1} \incmp^P_1 \{2_{2i-2}, 3_{2i-2}\}$ and that $\{2_{2i-2},3_{2i-2}\} <^P_1 \{2_{2i},3_{2i}\}$. It follows that $1_{2i-1} <^P_1 \{2_{2i},3_{2i}\}$ and that $\{2_{2i-2},3_{2i-2}\} <^P_1 \{2_{2i},3_{2i}\}$. From this, we can conclude that $\{2_{2i-2},3_{2i-2}\} <^P_1 1_{2i+1}$.

	Combining the above gives us the ordering $\{1_{2i-1},2_{2i-2},3_{2i-2}\} <^P_1 \mathbf{\{1_{2i+1},2_{2i},3_{2i}\}}$.
\end{enumerate}

Together, these four steps conclude our proof by induction.

\end{framed}

It follows that a prudent player will always simplify a list of simple values to a single simple one. They can not construct complex values. After all, almost every single pair of simple values is comparable, allowing the player to discard one of them. The only values incomparable with each other, from the perspective of player 1, are $2_i$, $3_i$ and $1_{i+1}$. $[2_i,3_i] = 1_{i+1}$ and any combination including $1_{i+1}$ can be simplified to $1_{i+1}$. This leads us to our final theorem:

\begin{framed}
\begin{theorem}
\label{theorem:prudentsimplicity}
All prudent short games with $N = 3$ and a given starting player result in a single simple value.
\end{theorem}

As mentioned in Section~\ref{sec:N-player games}, we assume the game to be converging. We can thus construct the full game tree. As we are given a fixed starting player, we can then determine for each node in the tree which player makes the corresponding choice. If we label the nodes in a bottom-up way, we can apply Theorem~\ref{theorem:prudentordering} to obtain a simple value in every node, as any combination of simple values, from the perspective of a given player, can be merged into another simple value.
\end{framed}

It might be interesting to note that the number of different values for a given starting player thus becomes at most linear in the size of the board. As each move, and thus each level in the game tree, removes a single token from the game and possibly isolates more tokens, the depth of the game tree --- and therefore also the exponent of a simple value --- can not exceed the number of initial tokens, which in turn can not exceed the number of vertices $n$ on the game board. As we have three different bases for simple values and at most $n$ different exponents ($0, \ldots, n-1$ as you need at least two tokens to make a move), this gives us at most $3n$ different values or outcome classes.

However, we have now fixed a starting player, so a position now consists of a configuration of the board and the player whose turn it is. Naturally, different values can be assigned to the same configuration depending on the starting player. For instance, the game
\begin{tabular}{|c|c|}
\hline
1 & 2 \\
\hline
\end{tabular}
is won either by player 1, if player 1 or player 3 starts, or by player 2, if player 2 starts. A different notation would be needed to construct a value that includes all possible starting players, as has been done for two-player games in classical combinatorial game theory.

\subsubsection*{Indifference}

Another possible approach, instead of playing prudently, would be to consider the values 2 and 3 to be equal from the perspective of player 1, so $2 =_1 3$. We call this an \emph{indifferent} player, as the player does not differentiate between the outcomes where they lose. They are simply losses, no matter which other player won. The assumption of an indifferent selfish player leads to the same ordering as in Theorem~\ref{theorem:prudentordering}, with two differences: the ordering uses the relation $<_1$ instead of $<^P_1$, and the values within sets of brackets are equal to each other instead of being incomparable. The proof is quite similar to the one in Theorem~\ref{theorem:prudentordering}, which we will leave as an exercise to the reader.

\section{Simplification results}
\label{sec:results}
In this section, we analyse the efficiency of our simplification rules by computing the number of different possible values for games of three-player Clobber on a $1 \times n$ board using the different simplification rules.

Recall Example~\ref{example:1x10-unsimplified} of a $1 \times 10$ board, which had a rather large value of 675 characters, excluding the spaces added for readability:

\begin{framed}
\textbf{Unsimplified: } [[[1, 3, [3, [[1, 3]]], [[1, 3, [2, 3]], [2, 3, [1, 2]]], [[[1, 2]]]], [3, [2, 3], [2, [1, 3]], [[1, 2]]], [3, [3, [1, 3]], [3, [[1, 3]]], [[1, 3], [2, 3]], [[1, 3]]], [3, [[1, 2], [2, 3]]], [[1, 3], [3, [1, 2], [[1, 2]]], [[1, 2, 3], [1, 2, [2, 3]]]], [[1, 3], [3, [1, 2], [[1, 2]]], [[1, 3], [2, 3, [2, 3]], [2, 3]]]], [[2, [1, 3], [2, 3], [[1, 3]]], [2, [3, [1, 3]]], [3, [1, [2, 3]], [[1, 3, [2, 3]]], [[1, 3], [[1, 2]]], [[2, 3], [[1, 3]]]], [[1, 3], [1, [2, 3]], [3, [2, 3]], [[2, 3]]], [[1, [1, 2, 3], [2, 3, [2, 3]]], [2, 3], [2, [[1, 2]]]], [[2, 3], [2, [1, 2]], [[1, 2, [2, 3]], [1, 2]]]], [[2, [2, 3]], [2, [3, [1, 3]], [3, [2, 3]]], [2, [3, [1, 3]]], [3, [2, 3], [[1, 3]], [[2, 3], [[1, 3]]]], [[1, 2], [1, 3, [1, 3]], [2, 3]], [[1, 2], [2, 3]]], [[[1, 2], [1, [2, 3]], [2, 3]], [[1, 2], [2, 3]], [[1, 3, [1, 2]]], [[2, 3]], [[3, [1, 3]], [[1, 3]]]]]
\end{framed}

Our syntactic rules turn out to be inapplicable in this case, but the assumption of three selfish players makes a huge difference, reducing the value to a simple one:

\begin{framed}
\textbf{Selfish: } [[[1,2]]].
\end{framed}

Note that this assumes that the first turn is player~1's. Three prudent players will get the same result, although they might express it as a simple value:

\begin{framed}
\textbf{Prudent: } $3_1$ (= [[[1,2]]]), brackets as seen from the perspective of player~1.
\end{framed}

To give a different example, where the assumption of prudent players simplifies the value more than just selfish players, consider the following game: \begin{tabular}{|c|c|c|c|c|c|c|c|c|}
\hline
1 & 3 & 2 & 3 & 2 & 3 & 1 & 2 & 3 \\
\hline
\end{tabular}. This gives us the following values:

\begin{framed}
\textbf{Selfish: } [[[1,2],[[2,3],[[1,3]]]],[[1,2],[[2,3]]],[[[2,3],[[1,3]]]]]

\textbf{Prudent: } $3_2$ (= [[[[2,3],[[1,3]]]]])
\end{framed}

Table~\ref{table:results} shows the number of unique possible values for boards of size $1~\times~n$ with $2 \leq n \leq 13$. We only analysed the configurations that do not occur on earlier board sizes. This means we skipped all configurations with a 0 at either extremity of the board or with at least two consecutive 0's, as these configurations would have already occurred at some smaller board size\footnote{Recall that a 0 is an empty vertex.}. We also took mirror symmetry into account, so we only analysed configurations whose string representation is lexicographically greater than or equal to their reverse's. Finally, following our assumption in Section~\ref{sec:N-player games}, we only considered configurations where at least one move is possible for some player. Because of this, the number of games we analysed is less than the number of actual possible configurations, which is $4^n$ for a $1 \times n$ board. Note that this also means that the number of values shown in the table is the number of different possible values at the starting position. Once the players proceed to make moves, different values may occur. For instance, with selfish players, some values occur on $1 \times 10$ boards that do not occur on $1 \times 11$ boards. Furthermore, we assume all players share the same preference --- we have not analysed a game in which, for instance, only one of the players is prudent --- and we assume that it is player 1's turn in each starting position, which seems to lower the number of possible values. On a $1 \times 4$ board with three prudent players, the value $1_1$ does occur while $2_1$ does not. Were player 2 the starting player, the value $2_1$ would have occurred, for instance on the board
\begin{tabular}{|c|c|c|c|}
\hline
1 & 2 & 2 & 3 \\
\hline
\end{tabular}. However, our results are from the perspective of the starting player, and we can always renumber the players so their number matches their turn order, to obtain a value from our results.

\begin{table}[ht!]
\begin{center}
\begin{tabular}{| r | r || r | r | r | r |}
\hline
Board length & Games analysed & Unsimplified & Syntactic & Selfish & Prudent \\
\hline
2 & 3 & 2 & 2 & 2 & 2 \\
3 & 15 & 3 & 3 & 3 & 3 \\
4 & 60 & 7 & 7 & 4 & 4 \\
5 & 243 & 21 & 21 & 5 & 5 \\
6 & 924 & 77 & 77 & 7 & 7 \\
7 & 3\,609 & 506 & 501 & 8 & 8 \\
8 & 13\,704 & 2\,408 & 2\,398 & 9 & 8 \\
9 & 52\,497 & 9\,777 & 9\,748 & 20 & 10 \\
10 & 199\,329 & 36\,407 & 36\,326 & 154 & 11 \\
11 & 758\,556 & 128\,345 & 128\,179 & 2\,163 & 13 \\
12 & 2\,878\,512 & 434\,571 & 434\,274 & 30\,378 & 13 \\
13 & 10\,949\,499 & 1\,441\,816 & 1\,441\,334 & 256\,975 & 14 \\
\hline
\end{tabular}
\caption{Number of Clobber games analysed and unique resulting values adding the different simplification methods. Note that the selfish and prudent columns also use the syntactic simplifications.}
\label{table:results}
\end{center}
\end{table}

As the table shows, the syntactic rules do reduce the number of unique values, but only very slightly. Selfish play reduces the numbers significantly on smaller board sizes, but the number of values still grows exponentially and the reduction factor seems to decrease as the board grows larger. This could be explained with the incomparability of the more complex values, as mentioned at the end of Section~\ref{sec:selfish}, as these more complex values occur more often on larger boards and a combination of two such values usually can not be simplified. As argued at the end of Section~\ref{sec:prudent}, prudent play results into a linear upper bound on the number of values. This is strenghtened by the results shown in the table, which also show that the upper bound of $3n$ is not sharp.

As a specific example, let us consider the single value that ``disappears'' when going from selfish to prudent play on a $1 \times 8$ board. This is [[1,3],[[1,2],[[2,3]]]], or, using the bar notation, $[\bar{2},\bar{\bar{2}}]$. This should indeed be simplified to just $\bar{2}$ by a prudent player 1. The other values occurring in selfish play are 1, 2, 3, $\bar{1}$, $\bar{2}$, $\bar{3}$, $\bar{\bar{1}}$ and $\bar{\bar{2}}$, which are all already simple values. As argued before, the value $\bar{\bar{3}}$ does not occur because we assume player 1 to be the starting player. We have also verified by hand that the reduction from 20 selfish to 10 prudent values for $1 \times 9$ is correct.

\section{Conclusions and further research}
\label{sec:conclusions}

In this paper, we have attempted to use simple player preferences to simplify the game tree in games with $N \geq 2$ players, and particularly with $N = 3$, using the game of Clobber as an example. We have presented two sets of generic player preferences which significantly reduce the number of unique values for arbitrary game positions --- our simplification rules for prudent play lead to a linear upper bound on this number for three-player games. These rules apply both to impartial and partisan games and are shown to work on Clobber. We postpruned our game trees, unfortunately meaning we still had to compute the entire tree before being able to simplify. While we have managed to significantly reduce the number of outcome classes for game values, our rules did not (significantly) lower the time needed to compute these values.

As we have only considered the outcome classes and not the victory margins, it is impossible to simply determine the value of a complex position from the values of its disjoint components. For instance, consider the games 
\begin{tabular}{|c|c|}
\hline
1 & 2 \\
\hline
\end{tabular}
, 
\begin{tabular}{|c|c|}
\hline
1 & 3 \\
\hline
\end{tabular}
 and 
\begin{tabular}{|c|c|c|}
\hline
1 & 1 & 2 \\
\hline
\end{tabular}
. These three games all have the value 1 if player 1 begins. Combining the first two, gives the game 
\begin{tabular}{|c|c|c|c|c|}
\hline
1 & 2 & 0 & 1 & 3 \\
\hline
\end{tabular}
, which has value $\bar{1}$ if player 1 begins, while combining the second two gives the game 
\begin{tabular}{|c|c|c|c|c|c|}
\hline
1 & 1 & 2 & 0 & 1 & 3 \\
\hline
\end{tabular}
, which has value $1$ if player 1 begins. While all components can be won by player 1 if they begin, player 1 can not begin in both games at once. Additional information is thus required to allow for a simple calculation of disjunctive sums, such as the options for the other players.

A logical next step would be to see how our simplification rules perform on games with more than three players. Furthermore, as our simple values are simply a means to simplify the notation of three-player games, it could be interesting to attempt to find a similar useful notation, and a generalisation of Theorem~\ref{theorem:prudentsimplicity}, for games with $N > 3$. A logical generalisation would be to keep the notation from Section~\ref{sec:simplevalues}: $a_i = \{b_{i-1}\,|\,b \neq a\}$. Using this notation, in a four-player game, we would have $2_1 = \{1, 3, 4\}$. However, there would be no simple notation for, for instance, $\{1, 4\}$, so it remains to be seen how useful this notation would be. Our final suggestion would be to devise more player preferences --- a risky or paranoid player, for instance --- and to mix several types of players to research the effects of different combinations. We have now assumed all players to have the same preference to experience the full effects of those specific rulesets, but naturally this is not always the case.


\end{document}